\documentclass[aps,showpacs,prl,floatfix,twocolumn]{revtex4-1}
\usepackage{bm}
\usepackage[dvipdfmx]{graphicx}
\usepackage{amsmath,amssymb}
\usepackage{float}
\usepackage{wrapfig}
\usepackage{layout}
\usepackage{color}
\usepackage[english]{babel}
\newcommand{\sss}{\mathrm{s}}
\newcommand{\ccc}{\mathrm{c}}
\newcommand{\ii}{\mathrm{i}}
\newcommand{\uu}{\mathrm{u}}
\newcommand{\cc}{\mathrm{c}}
\newcommand{\bb}{\mathrm{b}}

\newcommand{\dd}{\mathrm{d}}

\def\rightdef{=\mathrel{\mathop:}}

\usepackage{braket}
\begin{document}
\preprint{aps/}
\title{
Chiral solitons in monoaxial chiral magnets in tilted magnetic field
}

\author{Yusuke Masaki$^{1}$}
\email{masaki@vortex.c.u-tokyo.ac.jp}
\altaffiliation[Present address:]{Research and Education Center for Natural Science, Keio University, Yokohama, Kanagawa 223-8521, Japan}

\author{Ryuya Aoki$^{2}$}

\author{Yoshihiko Togawa$^{2}$}

\author{Yusuke Kato$^{1,3}$}
\affiliation{$^{1}$Department of Physics, The University of Tokyo, Bunkyo, Tokyo 113-0033, Japan}

\affiliation{$^{2}$Department of Physics and Electronics, Osaka Prefecture University, 1-1 Gakuencho, Sakai, Osaka 599-8531, Japan}

\affiliation{$^{3}$Department of Basic Science, The University of Tokyo, Meguro, Tokyo 153-8902, Japan}

\date{\today}

\begin{abstract}
We show that the stability (existence/absence) and interaction (repulsion/attraction) of chiral solitons in monoaxial chiral magnets can be varied by tilting the direction of magnetic field. We, thereby, elucidate that the condensation of attractive chiral solitons causes the discontinuous phase transition predicted by a mean field calculation. Furthermore we theoretically demonstrate that the metastable field-polarized-state destabilizes through the surface instability, which is equivalent to the vanishing surface barrier for penetration of the solitons. We experimentally measure the magnetoresistance (MR) of micrometer-sized samples in the tilted fields in demagnetization-free configuration. We corroborate the scenario that hysteresis in MR is a sign for existence of the solitons, through agreement between our theory and experiments.

\end{abstract}
\maketitle
{\it Introduction. }
Dzyaloshinskii--Moriya interactions (DMIs)\cite{Dzyaloshinsky1958,PhysRevLett.4.228,PhysRev.120.91} can exist in non-centrosymmetric magnets, where the competition between DMI and exchange interaction induces modulated spin structures\cite{Izyumov1984} such as conical, cycloidal and magnetic skyrmion states. Among those states, noteworthy are magnetic skyrmion lattice (SkL) in cubic chiral magnets\cite{BogdanovYablonskii1989,bogdanov1994thermodynamically,Muhlbauer2009,PhysRevLett.102.197202,NagaosaTokura2013} and chiral soliton lattice (CSL) in monoaxial chiral magnets\cite{Kishine2005,Kousaka2009,PhysRevLett.108.107202,Togawa2013,Togawa2016}; those two states consist of topological objects: single skyrmion in SkL and single discommensuration\cite{McMillan1976} (called chiral soliton in this paper) in CSL, respectively. Topological stability of those objects allows us to regard them as {\it emergent particles}. Their stability and interaction properties can be varied by elevating temperatures\cite{Schaub1985,Leonov2010,Leonov2018}. 
This controllability gives them an advantage in devise application in future spintronics. It is thus important to find more efficient way to control the physical characters of skyrmions and chiral solitons. In this paper, we show that tilting of the direction of magnetic field can change the interacting properties between repulsion and attraction, and stability/instability of chiral soliton in monoaxial chiral helimagnets, without utilizing temperature effects. We also show that interaction properties and stability/instability of chiral soliton account for the structure of the phase diagram at zero temperature found in an early mean field theory\cite{Laliena2016a}. Further we conduct magnetoresistance(MR) experiments for micrometer-sized samples of Cr$_{1/3}$NbS$_{2}$ in demagnetization-free configuration. We corroborate our theory on stability/instability of chiral soliton through quantitative agreement between the theory and the experiments. 

\begin{figure}
\includegraphics[width=25em]{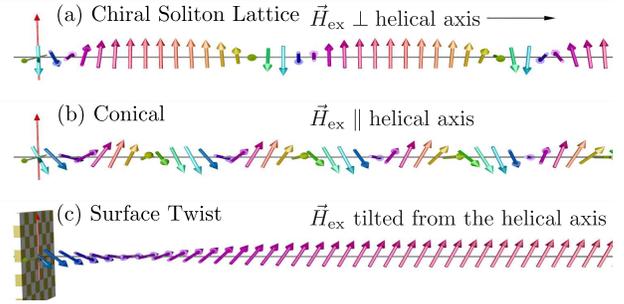}
\caption{Sketches of (a) chiral soliton lattice state, (b)conical state, (c) surface twist structure of a uniform state in the bulk. The helical axis is shown by the black arrow in (a). Colors of arrows stand for their directions in the plane perpendicular to the helical axis.}
\label{fig:sketch}
\end{figure}
{\it Monoaxial chiral magnets\footnote{
This terminology follows from recent references \cite{Togawa2016,Laliena2016a} for magnets with a helical propagation in one direction in contrast to the cubic chiral magnets.
In terms of symmetry, the DMIs of such a system are allowed to have $D_{\perp}\neq D_{\parallel}$, where $\perp$ and $\parallel$ are defined with regard to the direction of the helical propagation. 
We consider the case of $D_{\perp} = 0$ and write $D_{\parallel} = D$ in the following model. 
}. }
Cr$_{1/3}$NbS$_{2}$ is a monoaxial chiral magnet\cite{moriya1982evidence,MiyadaiKikuchiKondoSakkaAraiIshikawa1983,KishineOvchinnikov2015,Togawa2016}. It shows a helical state with its pitch of 48 nm along the $c$-axis, 
which we call the helical axis, 
in the absence of magnetic field\cite{moriya1982evidence,MiyadaiKikuchiKondoSakkaAraiIshikawa1983}. 
The helical structure consisting of spins rotating in the $ab$-plane is robust because of the strong hard-axis anisotropy along the helical axis. 
The magnetic field perpendicular to the helical axis induces an ideal chiral soliton lattice[Fig.~\ref{fig:sketch}(a)], and leads to a continuous phase transition (CPT) to the uniform state\cite{dzyaloshinskii1965theory_2}. Properties of Cr$_{1/3}$NbS$_{2}$ in equilibrium\cite{PhysRevLett.108.107202} and metastable states\cite{Togawa2015} have been quantitatively explained by Refs.~\cite{dzyaloshinskii1965theory_2} and \cite{Shinozaki2018}, respectively, with use of the chiral sine-Gordon model. Thus Cr$_{1/3}$NbS$_{2}$ is regarded as a {\it model material} of monoaxial chiral magnets. 
The field parallel to the helical axis induces a CPT from chiral conical state[Fig.~\ref{fig:sketch}(b)], to the uniform state.

Recently, Laliena {\it et al.}\cite{Laliena2016a} have found three types of field-induced phase transitions, which depend on the direction of magnetic field in monoaxial chiral magnets: CPT for fields with angle $\theta_H$ (with respect to the $ab$-plane) larger than $88.5^{\circ}$, discontinuous phase transition (DPT) for $81.5^{\circ}\lesssim \theta_H\lesssim 88.5^{\circ}$, another CPT for $0\le\theta_H\lesssim81.5^{\circ}$, and two multicritical points. In a subsequent paper\cite{Laliena2017}, they identified the former CPT as the instability-type and the latter CPT as the nucleation-type, following de Gennes's classification\cite{DeGennes1975}. The period of modulation in the ordered phase {\it diverges} in nucleation-type CPT, while a mode with a {\it finite} wave vector drives the instability-type CPT.  
Yonemura {\it et al.}\cite{Yonemura2017} recently performed field-sweep experiments of MR and magnetic torque measurements for micrometer-sized samples of Cr$_{1/3}$NbS$_{2}$ for various angles $\theta_H$ of magnetic fields. They found hysteresis loops and discrete steps for the angle $\theta_{H}\lesssim 88^{\circ}$, which includes the nucleation-type CPT as well as DPT, and  regarded them as evidences for chiral solitons. 

These results in Refs.~\cite{Laliena2016a, Yonemura2017} imply that the properties of chiral solitons in monoaxial chiral magnets depend on the direction of magnetic field. However, no explicit theoretical study along this direction has been done. Further, the origins of the DPT in the monoaxial chiral magnets found in Ref.~\cite{Laliena2016a} remain unclear so far. There has been no theory that supports quantitatively the arguments in Ref.~\cite{Yonemura2017}. In this paper, we address these issues.

\begin{figure}[t!]
\begin{center}
\includegraphics[width = 23em]{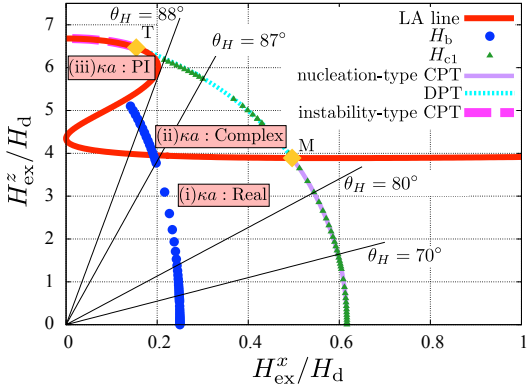}
\caption{
Phase diagram in the tilted magnetic field using realistic parameters shown in the text.
The solid red line is obtained using the linear analysis (LA line), which divides the phase diagram into three regions (i)--(iii). There are, correspondingly, three kinds of phase transition: nucleation-type CPT, the DPT, and the instability-type CPT denoted by the solid purple line, the dotted light-blue line, and the dashed pink line, respectively. Two yellow squares ``M'' and ``T'' represent multicritical and tricritical points respectively.
The low-field (high-field) side of the phase boundary is the ordered (disordered) phase. The phase boundary is obtained by minimizing the energy functional and basically the same as in Ref.~\cite{Laliena2016a}.
 The black solid lines with values of $\theta_{H}\equiv\tan^{-1}(H_{\mathrm{ex}}^{z}/H_{\mathrm{ex}}^{x})$ are guides to see the field angle.  
Solid circles labeled ``$H_{\bb}$'' and solid triangles labeled ``$H_{\cc1}$'', respectively, stand for the barrier field and the nucleation field defined in the text. }
\label{fig-phase-diagram}
\end{center}
\end{figure}

{\it Model.} 
We start with the following energy functional for the classical spins defined on a one dimensional lattice along the helical axis at zero temperature:
\begin{eqnarray}
E[\{\vec{M}_{l}\}]
&=& -\sum_{l} \Biggl[
	J_{\parallel} \vec{M}_{l}\cdot\vec{M}_{l+1} 
	+ D \left(\vec{M}_{l}\times \vec{M}_{l+1}\right)^{z} 
	\nonumber \\
&&\hspace{6em}	
	- \dfrac{K}{2} \left(M_{l}^{z}\right)^{2} 
	+\vec{H}_{\mathrm{ex}}\cdot \vec{M}_{l}
\Biggr]. \label{eq-energy-chain}
\end{eqnarray}
The local magnetic moment at site $l$ on the chain is given by $\vec{M}_{l}$, and the magnitude of each moment, $|\vec{M}_{l}|^{2}$, is 1.
The first and second terms are the Heisenberg exchange, and Dzyaloshinskii--Moriya interactions on the nearest neighbor pairs, respectively. The third term stands for the hard axis anisotropy for positive $K$. The last term is the Zeeman energy due to tilted magnetic field, $\vec{H}_{\mathrm{ex}}$, which has $x$- and $z$-components. As realistic parameters, we set $D = 0.16 J_{\parallel}$ and $K = 5.68 H_{\dd}$ with $H_{\dd} = 2[(J_{\parallel}^{2} + D^{2})^{1/2} - J_{\parallel}]$.  
The stationary condition is given by $\vec{M}_{l} \times \vec{H}_{l}^{\mathrm{eff}} = 0$ with
\begin{eqnarray}
\vec{M}_{l} &=& \hat{H}_{l}^{\mathrm{eff}} = \vec{H}_{l}^{\mathrm{eff}} / |\vec{H}_{l}^{\mathrm{eff}}|,\label{eq-moment}\\
\vec{H}_{l}^{\mathrm{eff}} & =&J_{\parallel} (\vec{M}_{l-1} +\vec{M}_{l+1}) +D\hat{z}\times (\vec{M}_{l-1}-\vec{M}_{l+1}) \nonumber\\
&&\hspace{11em}- KM_{l}^{z}\hat{z} + \vec{H}_{\mathrm{ex}}. \label{eq-field}
\end{eqnarray} 
{\it Properties of chiral solitons.}
We classify the region in the $H_{\mathrm{ex}}^{z}$-$H_{\mathrm{ex}}^{x}$ phase diagram, according to existence/absence  and interaction properties of chiral soliton, following the method used in  Ref.~\cite{Schaub1985}. 
Let us consider an isolated soliton with its center at $l=0$ and assume the following asymptotic form of the magnetic moment at $l \gg 1$:
\begin{eqnarray}
\vec{M}_{l}
=\vec{M}_{\uu} + \mathrm{Re}[\vec{A}\exp(-\kappa x_{l})]~\text{with}~x_{l} = l a, \label{eq-asymptotic-form}
\end{eqnarray}
where $\vec{M}_{\mathrm{u}}$ is the uniform solution without boundaries and $a$ is a lattice constant. A positive real part of $\kappa$ describes the soliton tail and corresponds to the inverse of the soliton size. On the other hand, a pure imaginary (PI) $\kappa = \ii q$, describes a distorted conical order with a fundamental wave number $q$ rather than an isolated soliton. 
In this case, the form \eqref{eq-asymptotic-form} is available for all $l$ when $\vec{A}$ is regarded as vanishingly small.  

Linearization of Eqs.~\eqref{eq-moment} and \eqref{eq-field} with respect to the second term of Eq.~(\ref{eq-asymptotic-form}) leads to the linear coupled-equation of $\vec{A}$ with condition $\vec{M}_{\uu} \cdot \vec{A} = 0$ deduced from the normalization, 
and we obtain the quadratic equation in $\cosh (\kappa a)$ through the condition for the existence of a non-trivial solution of $\vec{A}$. 
The values of $\kappa a $ depend on $H_{\mathrm{ex}}^{z}$ and $H_{\mathrm{ex}}^{x}$, and are classified into three cases through the discriminant: (i) real, (ii) complex, and (iii) PI. On the basis of the type of $\kappa a$, we draw the bold red line (``LA line'') in Fig.~\ref{fig-phase-diagram}, which separates the phase diagram into the three regions (i), (ii), and (iii). 
A necessary condition for the existence of an isolated soliton is that $\kappa a $ belongs to (i) or (ii), and actually there are instability lines of an isolated soliton in this region, which give the sufficient condition. 

Following Refs.~\cite{Jacobs1980,Schaub1985}, we summarize the interaction properties in the asymptotic region.
In the region (i), the interaction is repulsive for any inter-soliton distance. On the other hand, in the region (ii), the interaction energy oscillates as a function of the distance and can be attractive for some values of the distance. 
In Fig.~\ref{fig-phase-diagram}, we see that the interaction between solitons changes from repulsive to attractive in increasing $H_{\mathrm{ex}}^{z}$, the parallel component of the magnetic field.

{\it Comparison with the ground state phase diagram.}
In Fig.~2, we have also drawn the phase boundaries given by the three kinds of phase transitions. We see that the multicritical point M connecting the DPT line to the nucleation-type CPT line is located on the boundary between (i) and (ii).
In the region (i), the repulsion leads to a logarithmically diverging period near the transition. This explains the reason the phase transition in (i) is identified as nucleation-type CPT. On the other hand, in the region (ii), the 
attraction favors the periodic structure of solitons with a finite distance even at the transition point and leads to the DPT.  

The instability-type CPT around $(H_{\mathrm{ex}}^{x},H_{\mathrm{ex}}^{z}) \approx (0, H_{\cc}^{z})$ can be described as the development of a conical order with distortion owing to finite $H_{\mathrm{ex}}^{x}$, which is equivalent to the existence of vanishingly small $\vec{A}$ in the region (iii). 
A part of the LA line where (ii) and (iii) meet is the instability-type CPT line. The tricritical point T is located at the point where the LA line deviates from the phase boundary.

Consistency between properties of chiral soliton (the existence/absence and repulsion/attraction) and the types of phase transitions in the ground state phase diagram has two-fold implications; It explains the mechanisms of the phase transitions, and it endorses our arguments on properties of chiral solitons.

Recently, attractively interacting skyrmions in the conical phase, which result from a different mechanism, were theoretically studied\cite{Leonov2016} and experimentally confirmed by observing their clusters\cite{Loudon2018}. 
Attractive interaction between chiral solitons can be confirmed, in a similar way, i.e. by observation of the cluster formation of solitons in the uniform state.

\begin{figure}[t!]
\begin{center}
\includegraphics[width = 25em]{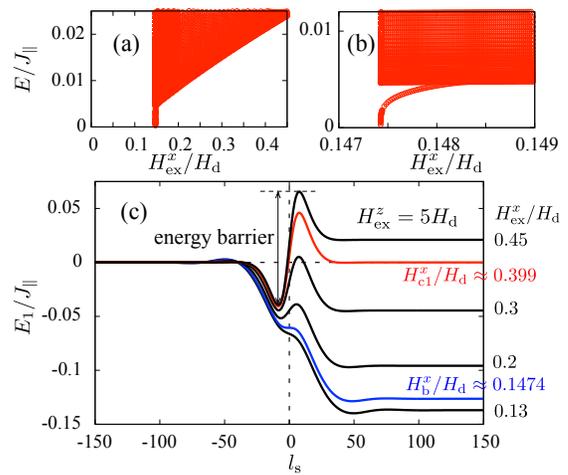}
\caption{
(a)~Excitation spectrum for the uniform state with surface twist. (b)~The magnified image of (a). The lowest eigenenergy becomes zero at around $H_{\mathrm{ex}}^{x} \approx 0.1474 H_{\dd}$. 
(c)~Energy landscapes of the isolated solitons for several values of $H_{\mathrm{ex}}^{x}$ indicated in the figure and  $H_{\mathrm{ex}}^{z} = 5H_{\dd}$. The horizontal axis represents the position of the soliton center, $l_{\mathrm{s}}$. 
$H_{\cc1}^{x}$ (red line) and $H_{\bb}^{x}$ (blue line) are the $x$-components of the nucleation field and the barrier field, respectively, when $H_{\mathrm{ex}}^{z} =5 H_{\dd}$.
}
\label{fig-surface-barrier}
\end{center}
\end{figure}
{\it Surface instability, surface barrier and hysteresis. }
So far we have seen that chiral solitons exist in the wide region of the phase diagram. Next we consider hysteresis observed in experiments for micrometer samples\cite{Togawa2015,Yonemura2017}. Particularly, the reproducible large jump in decreasing field is discussed in connection with surface instability and surface barrier for penetration of chiral solitons. 

First we perform the mode analysis in a way similar to that in Ref.~\cite{Muller2016}.
The detail is written in the supplemental material. Let us consider the field polarized state with surface twist\cite{Du2013,Du2013a,IwasakiMochizukiNagaosa2013a,Rohart2013, Sampaio2013, Wilson2013,Meynell2014,PhysRevB.89.224408} as a static configuration $\{\vec{M}_{\mathrm{s},l}\}$. Its structure is schematically shown in Fig.~\ref{fig:sketch}(c). The system is defined for $l\ge0$ with the free boundary condition $\vec{M}_{l=-1} = \vec{0}$, and thereby a twisted structure appears around the boundary. We obtain the excitation spectra from the equation of motion based on the bilinear form of energy~\eqref{eq-energy-chain} with respect to the normal modes for $\{\vec{M}_{\mathrm{s},l}\}$. The spectra are shown 
for $H_{\mathrm{ex}}^{z} = 5 H_{\dd}$ in Fig.~\ref{fig-surface-barrier}(a). The low energy state appears from the continuum spectra in decreasing field, as shown in Fig.~\ref{fig-surface-barrier}(b). This excitation is bound to the surface, leading to the penetration of the soliton. The energy becomes zero at $H_{\mathrm{ex}}^{x} \approx 0.1474H_{\dd}$, which is a surface-instability field\cite{Muller2016}.
Note that such a localized state is not always the destabilizing mode. Near the PI region, the lowest energy excitation leads to an instability of a conical order\footnote{See the supplemental material for the further detail.}.

Then we confirm that this instability field coincides with the field in which the surface barrier vanishes\cite{Shinozaki2018}.
Figure~\ref{fig-surface-barrier}(c) shows that the energy landscapes of an isolated chiral soliton as a function of the soliton center, $l_{\mathrm{s}}$, for several values of $H_{\mathrm{ex}}^{x}$ and  $H_{\mathrm{ex}}^{z} = 5H_{\dd}$.
Here the single soliton energy $E_{1}$ is measured from the uniformly polarized state\footnote{We define $E_{1} = E[\{\vec{M}_{l_{\mathrm{s}},l}\}] - E[\{\vec{M}_{\uu}\}]$, where the summation in $E[\{\vec{M}_{l}\}]$ is over $l\ge 0$ and  $\vec{M}_{l_{\mathrm{s}},l}$ is a spin profile obtained by arranging the single soliton solution to Eqs.~\eqref{eq-moment} and \eqref{eq-field} under the periodic boundary condition of a finite-size chain with its center at $l_{\mathrm{s}}\in \mathbb{Z}$. Note that the surface modulation is necessary for the genuine solution when the surface exists and the free boundary condition is imposed.}. This kind of energy landscape has been presented in Refs.~\cite{Bean1964,DeGennes1966} for a superconducting vortex, and in Refs.~\cite{IwasakiMochizukiNagaosa2013a,Shinozaki2018} for a chiral soliton. 
As is known in Ref.~\cite{Shinozaki2018}, there exist the characteristic local maximum and minimum structures inside and outside the system, respectively, for $H_{\mathrm{ex}}^{x} > H_{\bb}^{x}$. The surface barrier is described by the local maximum\cite{IwasakiMochizukiNagaosa2013a} while the surface twist is by the spin structure of an isolated soliton at the point of the local minimum\cite{Du2013,Du2013a,IwasakiMochizukiNagaosa2013a,Rohart2013, Sampaio2013, Wilson2013,Meynell2014,PhysRevB.89.224408}. They merge at $H_{\mathrm{ex}}^{x} = H_{\bb}^{x}$, i.e.,
the soliton outside the system for $H_{\mathrm{ex}}^{x} > H_{\bb}^{x}$ comes to the surface at $H_{\mathrm{ex}}^{x} = H_{\bb}^{x}$, and the surface barrier vanishes. Figure~\ref{fig-surface-barrier} shows that $H_{\mathrm{b}}^{x}$ is consistent with the instability field.

For these values of the tilted field, $\kappa a$ is complex, and correspondingly the interaction between solitons can be attractive in contrast to Refs.~\cite{Bean1964,Shinozaki2018}.
In this case, solitons are possibly {\it attracted to the surface}. 
Inside the system as well as outside, the energy landscape has local minima coming from the oscillation of the soliton profile. 
Particularly, the local minimum closest to the surface gives the global minimum inside the system.
A sufficiently small field step allows a few solitons to penetrate and be bound to local minima near the surface. This state might be observed by local measurements.
\begin{figure}[t] 
\begin{center}
\includegraphics[width = 26em]{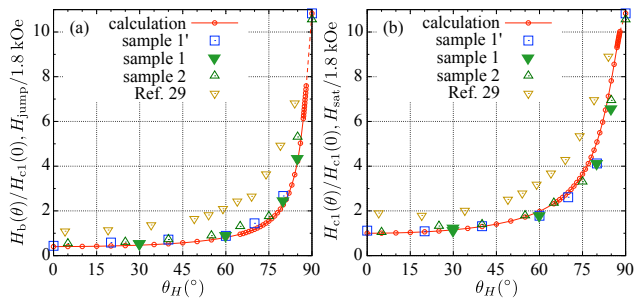}
\caption{Comparison between calculation and experiment: $H_{\bb}$ and $H_{\mathrm{jump}}$ (a) and $H_{\mathrm{c}1}$ and $H_{\mathrm{sat}}$ (b). The horizontal axes represent the angle of the tilted field and the vertical axes do the magnetic field. }
\label{fig-hbandhc}
\end{center}
\end{figure}

We calculate the barrier field, $\vec{H}_{\bb}$, in the region where solitons exist, as shown by blue solid circles in Fig.~\ref{fig-phase-diagram}, and directly compare the calculated values of $\vec{H}_{\bb}$ with experimentally observed jump fields below. In the PI region of  $\kappa a$, solitons do not exist, and the hysteresis is hardly observed in a magnetization process passing through this region.

{\it Magnetoresistance measurements. }
For quantitative comparison, we have to take account of the demagnetizing effects, which give difference between the internal and external fields.
We, thus, performed MR measurements in the configuration so as to avoid the demagnetizing effects. 
Dimensions of samples 1 and 2 are ($11.25 \mathrm{\mu m} * 0.7 \mathrm{\mu m} * 17.5 \mathrm{\mu m}$), and ($8.5 \mathrm{\mu m} * 0.5 \mathrm{\mu m} * 21 \mathrm{\mu m}$), respectively, where the order of the directions is $x*y*z(c\text{-axis})$. 
We define the tilted angle of the field $\theta_{H} = \tan^{-1}H_{\mathrm{ex}}^{z}/H_{\mathrm{ex}}^{x}$.
For samples 1 and 2, the field is in the plane of the film for 
any $\theta_{H}$, and demagnetizing effects on the field polarized states are small. 
The data taken from Ref.~\cite{Yonemura2017}, in which the sample dimension is ($0.7 \mathrm{\mu m} * 10 \mathrm{\mu m} * 17.5 \mathrm{\mu m}$) and it has large demagnetizing effects for $\theta_{H}\sim 0^{\circ}$, is shown for reference.

We performed two different sequences for sample 1, and label them sample 1 and sample 1'. 
The robustness of the hysteresis loops is confirmed through the multiple field-sweeps, 
where one sweep stands for a set of increasing and decreasing field processes. 
Actually five-time sweeps are done at $\theta_{H} = 30^{\circ}$, $60^{\circ}$, and $80^{\circ}$ for sample 1, 
and three-time sweeps are done at $\theta_{H} = 0^{\circ}$ for  sample 1', 
though only one sweep is done in the other cases\footnote{
See the supplementary material for experimental details and raw data of hysteresis loops.}.  
There are experimentally important two fields:  the saturation field $H_{\mathrm{sat}}$, where the hysteresis of MR closes in increasing field and the jump field $H_{\mathrm{jump}}$, where MR shows the sharp jump in decreasing field\cite{Togawa2015,Yonemura2017}.
We identify $H_{\mathrm{sat}}$ and $H_{\mathrm{jump}}$ as the theoretically important two fields, $H_{\cc}$ and $H_{\bb}$, respectively. Note that we use $H_{\cc 1}$, which is the nucleation field and defined so that the single soliton energy is zero, instead of $H_{\cc}$. For the nucleation-type CPT, $H_{\cc 1}$ is the same as $H_{\cc}$, while for the DPT, $H_{\cc 1}$ is slightly lower than $H_{\cc}$, but the difference is negligible as inferred  from Fig.~\ref{fig-phase-diagram}. 

We compare $H_{\mathrm{jump}}$ with $H_{\bb}$ in Fig.~\ref{fig-hbandhc}(a) and do  $H_{\mathrm{sat}}$ with  $H_{\cc1}$ in Fig.~\ref{fig-hbandhc}(b). $H_{\bb}$ and $H_{\cc1}$ are normalized by $H_{\mathrm{c}1} (\theta_{H} = 0^{\circ})$, while $H_{\mathrm{jump}}$ and $H_{\mathrm{sat}}$ are normalized by 1.8 kOe, which is the thermodynamic critical field at $\theta_{H}=0^{\circ}$ obtained in an experiment\cite{Yonemura2017}.
The value of the anisotropy is taken so that the critical field at $\theta_{H} = 90^{\circ}$ is $19.5~\mathrm{kOe}$.

The angle dependences of $H_{\bb}$ and $H_{\cc1}$ agree well with those of $H_{\mathrm{jump}}$ and $H_{\mathrm{sat}}$, respectively,  as shown in Figs.~\ref{fig-hbandhc}(a) and (b), except for the data of Ref.~\cite{Yonemura2017}, in which disagreement is caused by large demagnetizing effects. 
This consistency for the whole range of the phase diagram strongly supports the scenario for the clear hysteresis. The hysteresis due to the surface effects does not conflict with the type of phase transitions discussed in the ground state phase diagram. 
Agreement can be improved by taking into account the demagnetizing effects, but our approach sufficiently explains the physical origin of the characteristic hysteresis as a starting point. 

{\it Discussion. }
Earlier studies\cite{Schaub1985,Leonov2011a,Leonov2018} have discussed attractive interaction between solitons/skyrmions due to ``soft modulus effects''\cite{Leonov2011a}, i.e. effects due to spatial variation of modulus of local magnetic moment. This effect becomes important at finite temperatures, although they have not been experimentally confirmed yet. Our study demonstrates that soft modulus effects exist even at {\it zero temperature} by tilting magnetic field; Reduction of the in-plane moduli of local magnetic moments can change spin profiles, interaction properties and stability of chiral solitons. At zero temperature, whether soft modulus effects are possible depends on the manifold of topological defects. The soliton is a defect of in-plane components (XY spins) and has an extra direction for softening of in-plane-amplitude, while the skyrmion is that of Heisenberg spins and thus does not have soft modulus effects due to this mechanism.

As another origin of soft modulus effects, quantum fluctuation is worthwhile to consider in future study. Thermal fluctuation, quantum fluctuation, tilting magnetic field and their combination will open a wider possibility to control the physical properties of chiral soliton in chiral magnets. 

\begin{acknowledgements}
Y.~M. thanks H.~Tsunetsugu and J.~Kishine for helpful comments. 
Y.~K. and Y.~M. thank Alex Bogdanov for his introduction of nucleation-type phase transition during his stay in Komaba in Tokyo in early 2017. We acknowledge support under Japan Society for the Promotion of Science (JSPS) KAKENHI Grants No.~16J03224, No.~17H02923, No.~17H02767, and No.~25220803. This work was also supported by Chirality Research Center (Crescent) in Hiroshima University, the Mext program for promoting the enhancement of research universities, Japan, JSPS, Russian Foundation for Basic Research (RFBR) under the Japan - Russian Research Cooperative Program, and JSPS Core-to-Core Program, A. Advanced Research Networks, and
the Program for Leading Graduate Schools, the Ministry of Education, Culture, Sports, Science and Technology, Japan.
\end{acknowledgements}

\bibliographystyle{apsrev4-1}

\bibliography{library}

\clearpage
\pagebreak
\widetext
\begin{center}
\textbf{\large Supplemental Materials: Chiral soliton in monoaxial chiral magnets under tilted magnetic field}
\end{center}
\setcounter{equation}{0}
\setcounter{figure}{0}
\setcounter{table}{0}
\setcounter{page}{1}
\makeatletter
\renewcommand{\theequation}{S\arabic{equation}}
\renewcommand{\thefigure}{S\arabic{figure}}
\renewcommand{\bibnumfmt}[1]{[S#1]}
\renewcommand{\citenumfont}[1]{S#1}

\section{Experimental detail}
\begin{figure}[h!]
\begin{center}
\includegraphics[width = 160mm]{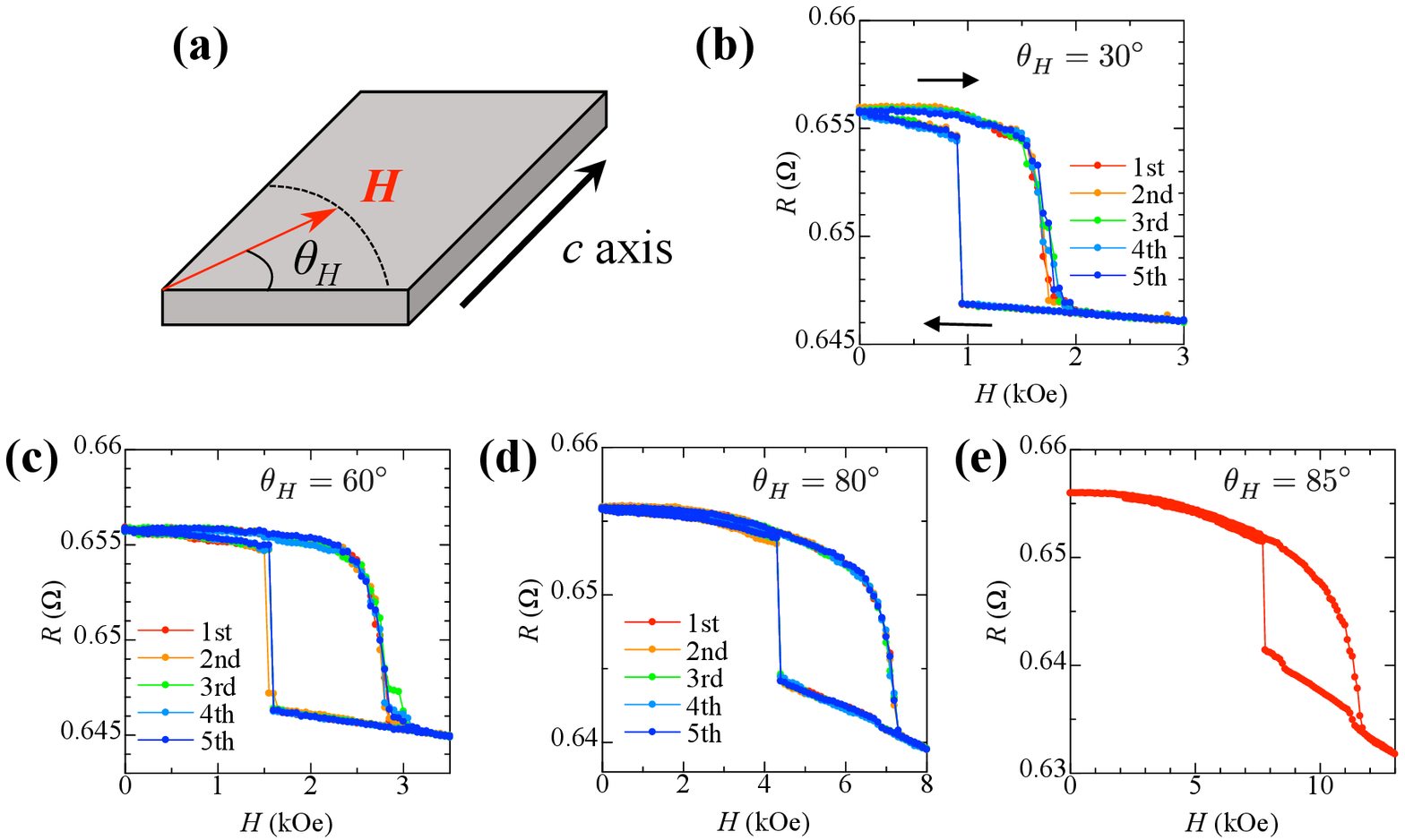}
\caption{(a) Schematic of the specimen and magnetic field configuration. (b)-(e) Magnetoresistance data in increasing and decreasing field processes for the sample~1 at 10~K. All data for five times field cycles are plotted in each panel except for 85~degree.}
\label{fig-MR}
\end{center}
\end{figure}
Bulk single crystals of CrNb$_{3}$S$_{6}$ were grown by chemical vapour transport method as described elsewhere~\cite{S-Kousaka2009}. Micrometer-sized platelet specimens were cut from the bulk single crystal used in Ref.~\cite{S-Yonemura2017} by using a focused ion beam (FIB) machine. Gold electrodes were prepared on the specimens for four-terminal resistance measurements by means of electron beam lithography (EBL) and lift-off techniques. The specimen dimensions are given in the main text. The resistance measurements were performed using a four-terminal method with ac current whose amplitude was 1.0~mA and frequency was 137~Hz. Magnetic field direction was rotated in the specimen plane to minimize the contribution of demagnetizing effect as schematically drawn in Fig. S1(a). The angle is defined as 0~degree when $H$ is perpendicular to the $c$ axis of the specimen, while 90~degree in the configuration with $H$ parallel to the $c$ axis. Figures S1(b) to S1(e) present the magnetoresistance data of the sample~1 at 10~K at 30, 60, 80, and 85~degrees, respectively. The measurements were performed five times except for the data at 85~degree. The magnetic field intervals are 50~Oe for the data taken at 30 and 60~degrees, and 100~Oe for at 80 and 85~degrees.

\section{Detail of mode analysis}
\subsection{Formulation}
We summarize the eigenequation for normal modes in the presence of the modulated structure as a static solution.
We start with the following Hamiltonian of the monoaxial chiral magnets
\begin{align}
\mathcal{H}&=-\sum_{\bm{j}}
\left[
 J_{\parallel} \vec{M}_{\bm{j}}\cdot \vec{M}_{\bm{j} + \hat{z}}
+D\vec{e}^{\ \!\!\ \!\! z} \cdot\left(\vec{M}_{\bm{j}}\times \vec{M}_{\bm{j} + \hat{z}}\right) + \vec{H}_{\mathrm{ex}} \cdot \vec{M}_{\bm{j}} - \dfrac{K}{2}\left(\vec{M}_{\bm{j}}\cdot \vec{e}^{\ \!\!\ \!\!z}\right)^{2}\right] +\sum_{\bm{j}} \mathcal{H}_{\perp,\bm{j}}, \\
\mathcal{H}_{\perp,\bm{j}} &=-\sum_{\hat{\mu}= \hat{x},\hat{y}} \left[J_{\mu} \vec{M}_{\bm{j}}\cdot \vec{M}_{\bm{j} + \hat{\mu}}
+D_{\mu}\vec{e}^{\ \!\!\ \!\! \mu} \cdot\left(\vec{M}_{\bm{j}}\times \vec{M}_{\bm{j} + \hat{\mu}}\right) 
\right].
\end{align}
Interactions between in-plane spins, $J_{\mu}$ and $D_{\mu}$ are independent of the direction $\mu=x,y$ in the case of the monoaxial magnet, and we can write them as $J_{x} = J_{y} = J_{\perp}$ and $D_{x} = D_{y} = D_{\perp}$.
Here we specify a site on a cubic lattice as $\bm{j}=\bm{j}_{\perp} + l \hat{z} = j\hat{x}+ k\hat{y} + l\hat{z}$. Let us consider the modulated structure in $z$-direction given by
\begin{align}
\vec{M}_{\sss,\bm{j}} = {}^{t}\!(\cos \varphi_{\sss,l}\sin \theta_{\sss,l},\sin \varphi_{\sss,l}\sin \theta_{\sss,l},\cos\theta_{\sss,l})
\end{align}
and new spin coordinate system given by
\begin{align}
\vec{M}_{\bm{j}} = \check{U}_{l}\vec{\tilde{M}}_{\bm{j}},~~
\check{U}_{l} = 
\begin{pmatrix}
-\sin \varphi_{\sss,l} & -\cos\varphi_{\sss,l}\cos\theta_{\sss,l} &\cos \varphi_{\sss,l}\sin \theta_{\sss,l} \\
\cos \varphi_{\sss,l} & -\sin\varphi_{\sss,l}\cos\theta_{\sss,l} & \sin \varphi_{\sss,l}\sin \theta_{\sss,l} \\
0 & \sin \theta_{\sss,l} & \cos\theta_{\sss,l}
\end{pmatrix}.
\end{align}
Subscripts $\sss$ denote {\it static}.
We introduce the unit vectors in the tilde frame as 
\begin{align}
\vec{\tilde{M}}_{\bm{j}} = \tilde{M}^{x}_{\bm{j}} \vec{\tilde{e}}_{l}^{\ \!\!\ \!\!x}
+ \tilde{M}^{y}_{\bm{j}} \vec{\tilde{e}}_{l}^{\ \!\!\ \!\!y} + \tilde{M}^{z}_{\bm{j}} \vec{\tilde{e}}_{l}^{\ \!\!\ \!\!z},
\end{align}
where
\begin{align}
\check{U}_{l} = (\vec{\tilde{e}}_{l}^{\ \!\!\ \!\!x},\vec{\tilde{e}}_{l}^{\ \!\!\ \!\!y},\vec{\tilde{e}}_{l}^{\ \!\!\ \!\!z}),~\text{and}~
\vec{\tilde{e}}_{l}^{\ \!\!\ \!\!x} \cdot \vec{e}^{\ \!\!\ \!\!z} = 0,~\vec{\tilde{e}}_{l}^{\ \!\!\ \!\!z} \cdot \vec{M}_{\sss,\bm{j}} = 1.
\end{align}
Introducing the following Fourier transform:
\begin{align}
\tilde{M}_{\bm{j}}^{x,y} = \dfrac{1}{\sqrt{N_{2\dd}}}\sum_{\bm{k}_{\perp}=k_{x},k_{y}} \tilde{M}_{\bm{k}_{\perp},l}^{x,y}e^{\ii \bm{k}_{\perp}\cdot \bm{j}_{\perp} a}
\end{align}
and we write down the Hamiltonian up to second order of $\tilde{M}_{l}^{x}$ and $\tilde{M}_{l}^{y}$ in the form 
\begin{align}
\mathcal{H} =E(\{\varphi_{\sss,l}\},\{\theta_{\sss,l}\}) + \dfrac{1}{2}\sum_{\bm{k}_{\perp}}\sum_{l,m}\sum_{\mu,\nu = x,y}
\tilde{M}_{-\bm{k}_{\perp},l}^{\mu} \mathcal{K}_{l,m}^{\mu,\nu}(\bm{k}_{\perp})\tilde{M}_{\bm{k}_{\perp},m}^{\nu}.
\end{align} 
It is obvious that $\mathcal{K}$ is hermitian in the sense that $(\mathcal{K}_{l,m}^{\mu,\nu}(\bm{k}_{\perp}))^{*}=\mathcal{K}_{l,m}^{\mu,\nu}(-\bm{k}_{\perp}) = \mathcal{K}_{m,l}^{\nu,\mu}(\bm{k}_{\perp})$. The first order terms of $\tilde{M}_{l}^{x}$ and $\tilde{M}_{l}^{y}$ vanishes owing to the equilibrium condition of $\varphi_{\sss,l}$ and $\theta_{\sss,l}$. 
Note that $\sum_{\bm{j}_{\perp}}\tilde{M}_{\bm{j}}^{z} \approx \sum_{\bm{k}_{\perp}}1 -[\tilde{M}_{-\bm{k}_{\perp},l}^{x}\tilde{M}_{\bm{k}_{\perp},l}^{x} + \tilde{M}_{-\bm{k}_{\perp},l}^{y}\tilde{M}_{\bm{k}_{\perp},l}^{y}]/2$.
\begin{description}
\item[Exchange interaction]~\\
For convenience, we use the notations $\cos \theta_{\sss,l}  \rightdef\ccc \theta_{\sss,l}$, $\sin \theta_{\sss,l}  \rightdef\sss \theta_{\sss,l}$, and $\varphi_{\sss,l} -\varphi_{\sss,l+1} \rightdef \Delta\varphi_{\sss,l}$.
The exchange term is transformed using $\tilde{M}$ as
$\vec{M}_{\bm{j}}\cdot\vec{M}_{\bm{j}+\hat{z}} = \sum_{\mu,\nu=x,y,z}\tilde{M}_{\bm{j}}^{\mu} \vec{\tilde{e}}_{l}^{\ \!\!\ \!\!\mu}\cdot \vec{\tilde{e}}_{l+1}^{\ \!\!\ \!\!\nu}\tilde{M}_{\bm{j}+\hat{z}}^{\mu} $
\begin{align}
\vec{\tilde{e}}_{l}^{\ \!\!\ \!\!\mu}\cdot \vec{\tilde{e}}_{l+1}^{\ \!\!\ \!\!\nu}
=
\begin{pmatrix}
\cos \Delta\varphi_{\sss,l}&
\ccc \theta_{\sss,l+1} \sin \Delta\varphi_{\sss,l} &
-\sss \theta_{\sss,l+1} \sin \Delta\varphi_{\sss,l}\\
-\ccc \theta_{\sss,l} \sin \Delta\varphi_{\sss,l} & 
\ccc \theta_{\sss,l}\ccc\theta_{\sss,l+1} \cos\Delta\varphi_{\sss,l}+\sss \theta_{\sss,l}\sss\theta_{\sss,l+1} &
-\ccc \theta_{\sss,l}\sss \theta_{\sss,l+1}\cos\Delta\varphi_{\sss,l} + \sss \theta_{\sss,l}\ccc \theta_{\sss,l+1} \\
\sss \theta_{\sss,l} \sin \Delta\varphi_{\sss,l} &
-\sss \theta_{\sss,l}\ccc \theta_{\sss,l+1}\cos \Delta \varphi_{\sss,l}+ \ccc\theta_{\sss,l}\sss \theta_{\sss,l+1} &
\sss \theta_{\sss,l}\sss\theta_{\sss,l+1} \cos\Delta\varphi_{\sss,l}+\ccc \theta_{\sss,l}\ccc\theta_{\sss,l+1}
\end{pmatrix}^{\mu\nu}.
\end{align}

\item[Dzyaloshinskii--Moriya interaction]~\\
The second term is written as $\vec{e}^{\ \!\!\ \!\!z}\cdot(\vec{M}_{\bm{j}}\times\vec{M}_{\bm{j}+\hat{z}}) 
= \sum_{\mu,\nu=x,y,z}\tilde{M}_{\bm{j}}^{\mu}
  [\vec{e}^{z}\cdot(\vec{\tilde{e}}_{l}^{\ \!\!\ \!\!\mu}\times
\vec{\tilde{e}}_{l+1}^{\ \!\!\ \!\!\nu})^{z}]\tilde{M}_{\bm{j}+\hat{z}}^{\nu}
$ and we calculate the matrix element as follows:
\begin{align}
 [(\vec{\tilde{e}}_{l}^{\ \!\!\ \!\!\mu}\times
\vec{\tilde{e}}_{l+1}^{\ \!\!\ \!\!\nu})^{z}]
=\begin{pmatrix}
-\sin\Delta\varphi_{\sss,l} & 
\ccc \theta_{\sss,l+1} \cos\Delta\varphi_{\sss,l}&
-\sss \theta_{\sss,l+1} \cos\Delta\varphi_{\sss,l}\\
-\ccc \theta_{\sss,l} \cos\Delta\varphi_{\sss,l}&
-\ccc\theta_{\sss,l}\ccc\theta_{\sss,l+1}\sin\Delta\varphi_{\sss,l} &
\ccc \theta_{\sss,l}\sss \theta_{\sss,l+1} \sin\Delta\varphi_{\sss,l}\\
\sss \theta_{\sss,l} \cos\Delta\varphi_{\sss,l}&
\sss \theta_{\sss,l}\ccc \theta_{\sss,l+1} \sin\Delta\varphi_{\sss,l}&
-\sss\theta_{\sss,l}\sss\theta_{\sss,l+1}\sin\Delta\varphi_{\sss,l} 
\end{pmatrix}^{\mu\nu}.
\end{align}
\item[Zeeman coupling]~\\
The third term is given by $\vec{H}_{\mathrm{ex}}\cdot \vec{M}_{\bm{j}} = \sum_{\mu}\vec{H}_{\mathrm{ex}}\cdot \vec{\tilde{e}}_{l}^{\ \!\!\ \!\!\mu}\tilde{M}_{\bm{j}}^{\mu}\to \vec{H}_{\mathrm{ex}}\cdot\vec{\tilde{e}}_{l}^{\ \!\!\ \!\!z} \tilde{M}_{\bm{j}}^{z}$. In the final transformation, we retain the term contributing the equilibrium state energy and the second order expansion. 
\begin{align}
 \vec{H}_{\mathrm{ex}}\cdot\vec{\tilde{e}}_{l}^{\ \!\!\ \!\!z}
 =H_{\mathrm{ex}}^{x}\cos\varphi_{\sss,l}\sss\theta_{\sss,l}
 +H_{\mathrm{ex}}^{z}\ccc\theta_{\sss,l}.
\end{align}
\item[Anisotropy]~\\
The fourth term is given by $(\vec{M}_{\bm{j}}\cdot \vec{e}^{\ \!\!\ \!\!z} )^{2}= \sum_{\mu,\nu=x,y,z}\tilde{M}_{\bm{j}}^{\mu} (\vec{\tilde{e}}_{l}^{\ \!\!\ \!\!\mu}\cdot\vec{e}^{\ \!\!\ \!\!z})(\vec{\tilde{e}}_{l}^{\ \!\!\ \!\!\nu}\cdot\vec{e}^{\ \!\!\ \!\!z})\tilde{M}_{\bm{j}}^{\nu}$.
\begin{align} 
(\vec{\tilde{e}}_{l}^{\ \!\!\ \!\!\mu}\cdot\vec{e}^{\ \!\!\ \!\!z})(\vec{\tilde{e}}_{l}^{\ \!\!\ \!\!\nu}\cdot\vec{e}^{\ \!\!\ \!\!z})
=\begin{pmatrix}
0 & 0 & 0\\
0 & 
\sss^{2} \theta_{\sss,l}&
\sss \theta_{\sss,l}\ccc \theta_{\sss,l}\\
0 &
\sss \theta_{\sss,l}\ccc \theta_{\sss,l}&
\ccc^{2} \theta_{\sss,l}
\end{pmatrix}.
\end{align}
\item[In-plane interactions]~\\
In-plane exchange and DMIs have dependence on the  in-plane wave vector. We consider the in-plane DMIs of the form
\begin{align}
-\sum_{\bm{j}}\sum_{\rho=x,y}D_{\rho}\left(\vec{M}_{\bm{j}}\times \vec{M}_{\bm{j} + \hat{\rho}}\right)\cdot \vec{e}^{\rho}.
\end{align}
We transform $\sum_{\bm{j}}\mathcal{H} _{\perp,\bm{j}}$ as $\frac{1}{2}\sum_{\bm{j}}\sum_{\mu,\nu=x,y,z}\sum_{\rho=x,y}\tilde{M}_{\bm{j}}^{\mu} \mathcal{K}_{\mathrm{int},\bm{j},\bm{j}+\hat{\rho}}^{\mu\nu}\tilde{M}_{\bm{j} + \hat{\rho}}^{\nu}$, and $\mathcal{K}_{\mathrm{int}}$ is given by
\begin{align}
\mathcal{K}_{\mathrm{int},\bm{j},\bm{j}+\hat{x}} = -2J_{x} \begin{pmatrix}
1 & 0 & 0 \\
0 & 1 & 0 \\
0 & 0 & 1 \end{pmatrix} -
2D_{x}
\begin{pmatrix}
0 & 
\cos \varphi_{\sss,l} \sss\theta_{\sss,l} &
\cos \varphi_{\sss,l} \ccc\theta_{\sss,l} \\
-\cos \varphi_{\sss,l} \sss\theta_{\sss,l} &
0&
-\sin \varphi_{\sss,l}\\
-\cos \varphi_{\sss,l} \ccc\theta_{\sss,l} &
\sin \varphi_{\sss,l}&
0
\end{pmatrix}
\\
\mathcal{K}_{\mathrm{int},\bm{j},\bm{j}+\hat{y}} = -2J_{y} \begin{pmatrix}
1 & 0 & 0 \\
0 & 1 & 0 \\
0 & 0 & 1 \end{pmatrix} -
2D_{x}
\begin{pmatrix}
0 & 
\sin \varphi_{\sss,l} \sss\theta_{\sss,l} &
\sin \varphi_{\sss,l} \ccc\theta_{\sss,l} \\
-\sin\varphi_{\sss,l} \sss\theta_{\sss,l} &
0&
\cos \varphi_{\sss,l}\\
-\sin \varphi_{\sss,l} \ccc\theta_{\sss,l} &
-\cos \varphi_{\sss,l}&
0
\end{pmatrix}.
\end{align}
Using $\tilde{M}_{\bm{j}}^{z} \simeq 1- \sum_{\mu=x,y}(\tilde{M}_{\bm{j}}^{\mu})^{2}/2$, $\sum_{\bm{j}}\mathcal{H} _{\perp,\bm{j}}$ is reduced to 
\begin{align}
\sum_{l,\bm{k}_{\perp}}\left\{
-(J_{x}+J_{y})+
\dfrac{1}{2}\begin{pmatrix}
\tilde{M}_{-\bm{k}_{\perp},l}^{x} &
\tilde{M}_{-\bm{k}_{\perp},l}^{y}
\end{pmatrix}
\left[
2\sum_{\mu = x,y}\begin{pmatrix}
J_{\mu}(1-\cos k_{\mu}a) & -\ii D_{\mu}M_{\sss,\bm{j}}^{\mu} \sin k_{\mu}a  \\
\ii D_{\mu}M_{\sss,\bm{j}}^{\mu}\sin k_{\mu}a  & J_{\mu} ( 1-\cos k_{\mu}a )
\end{pmatrix}
\right]
\begin{pmatrix}
\tilde{M}_{\bm{k}_{\perp},l}^{x} \\
\tilde{M}_{\bm{k}_{\perp},l}^{y}
\end{pmatrix}\right\}.
\end{align}

\end{description}
We summarize the above expressions.
Defining $\tilde{J} = \sqrt{J_{\parallel}^{2} + D^{2}}$ and $\tan\alpha = D/J_{\parallel}$, we  obtain the explicit forms of $E$ and $\mathcal{K}$ as follows:
\begin{align}
E(\{\varphi_{\sss,l}\},\{\theta_{\sss,l}\})
=-N_{2\dd}\sum_{l}&\Biggl\{\tilde{J}(
\sss \theta_{\sss,l}\sss\theta_{\sss,l+1} \cos(\Delta\varphi_{\sss,l} + \alpha)+J_{\parallel}\ccc \theta_{\sss,l}\ccc\theta_{\sss,l+1}
\nonumber\\
&\hspace{5em}
+H_{\mathrm{ex}}^{x}\cos\varphi_{\sss,l}\sss\theta_{\sss,l}
+H_{\mathrm{ex}}^{z}\ccc\theta_{\sss,l}
-\dfrac{K}{2}\ccc^{2}\theta_{\sss,l}
+ (J_{x}+J_{y})\Biggr\} 
\end{align}
and 
\begin{align}
\mathcal{K}_{l,l+1}^{xx}
&=-\tilde{J}\cos(\Delta\varphi_{\sss,l} + \alpha) \\
\mathcal{K}_{l,l+1}^{xy}
&=-\tilde{J}\ccc\theta_{\sss,l+1}\sin(\Delta\varphi_{\sss,l} + \alpha)\\
\mathcal{K}_{l,l+1}^{yx}
&=+\tilde{J}\ccc\theta_{\sss,l}\sin(\Delta\varphi_{\sss,l}+\alpha)\\
\mathcal{K}_{l,l+1}^{yy}
&=-\tilde{J}
\ccc \theta_{\sss,l}\ccc\theta_{\sss,l+1}\cos(\Delta\varphi_{\sss,l}+\alpha) -J_{\parallel}\sss \theta_{\sss,l}\sss\theta_{\sss,l+1}\\
\mathcal{K}_{l,l}^{xx}
&=
\tilde{J}
\sss \theta_{\sss,l}[\sss\theta_{\sss,l+1} \cos(\Delta\varphi_{\sss,l}+\alpha)
+\sss \theta_{\sss,l-1}\cos(\Delta\varphi_{\sss,l-1}+\alpha)]
+J_{\parallel}
\ccc \theta_{\sss,l}(\ccc\theta_{\sss,l+1}
+\ccc \theta_{\sss,l-1}) 
\nonumber \\
&+H_{\mathrm{ex}}^{x}\cos\varphi_{\sss,l}\sss\theta_{\sss,l}
+H_{\mathrm{ex}}^{z}\ccc\theta_{\sss,l} - K \ccc^{2}\theta_{\sss,l}+2\sum_{\mu=x,y}J_{\mu}(1-\cos k_{\mu}a)\\
\mathcal{K}_{l,l}^{xy}
&= -2\ii \sum_{\mu=x,y}D_{\mu}M_{\sss,\bm{j}}^{\mu}\sin k_{\mu}a\\
\mathcal{K}_{l,l}^{yx}
&= 2\ii \sum_{\mu=x,y} D_{\mu}M_{\sss,\bm{j}}^{\mu}\sin k_{\mu}a\\
\mathcal{K}_{l,l}^{yy}
&=
\tilde{J}
\sss \theta_{\sss,l}[\sss\theta_{\sss,l+1} \cos(\Delta\varphi_{\sss,l}+\alpha)
+\sss \theta_{\sss,l-1}\cos(\Delta\varphi_{\sss,l-1}+\alpha)]
+J_{\parallel}
\ccc \theta_{\sss,l}(\ccc\theta_{\sss,l+1}
+\ccc \theta_{\sss,l-1})
\nonumber \\
&+H_{\mathrm{ex}}^{x}\cos\varphi_{\sss,l}\sss\theta_{\sss,l}
+H_{\mathrm{ex}}^{z}\ccc\theta_{\sss,l} - K (\ccc^{2}\theta_{\sss,l}-\sss^{2}\theta_{\sss,l})
+2\sum_{\mu=x,y}J_{\mu}(1-\cos k_{\mu}a).
\end{align}
Note the relation $(\mathcal{K}_{l,m}^{\mu\nu}(\bm{k}_{\perp}))^{*}=\mathcal{K}_{l,m}^{\mu\nu}(-\bm{k}_{\perp}) = \mathcal{K}_{m,l}^{\nu\mu}(\bm{k}_{\perp})$, and the other components are zero. Our equation of motion is given by $
\frac{\dd \vec{M}_{\bm{j}}}{\dd t} = - \vec{M}_{\bm{j}} \times \left(-\frac{\partial \mathcal{H}}{\partial \vec{M}_{\bm{j}}}\right)$, which now reads
\begin{align}
-\ii \omega 
\begin{pmatrix}
\tilde{M}_{\bm{k}_{\perp},l}^{x} \\
\tilde{M}_{\bm{k}_{\perp},l}^{y}
\end{pmatrix}
= 
\begin{pmatrix}
-(\partial \mathcal{H})/(\partial\tilde{M}_{-\bm{k}_{\perp},l}^{y})\\
(\partial \mathcal{H})/(\partial\tilde{M}_{-\bm{k}_{\perp},l}^{x})
\end{pmatrix}
=\sum_{m}\begin{pmatrix}
-\mathcal{K}_{l,m}^{yx}(\bm{k}_{\perp}) &
-\mathcal{K}_{l,m}^{yy}(\bm{k}_{\perp}) \\
\mathcal{K}_{l,m}^{xx}(\bm{k}_{\perp}) &
\mathcal{K}_{l,m}^{xy}(\bm{k}_{\perp})
\end{pmatrix}
\begin{pmatrix}
\tilde{M}_{\bm{k}_{\perp},m}^{x}\\
\tilde{M}_{\bm{k}_{\perp},m}^{y}\\
\end{pmatrix}. \label{eq-bog}
\end{align}

\subsection{Numerical scheme}
We numerically diagonalize Eq.~\eqref{eq-bog} to obtain the excitation spectra and eigenvectors using a software of CPPlapack. 
We consider the sufficiently large finite-size lattice chain, the number of the site in $z$-direction, $N_{z}$, is set to 2000 ($l = 0, \cdots, N_{z}-1$). The free boundary condition is given by $\vec{M}_{l=-1} = \vec{0}$ and $\vec{M}_{l=N_{z}} = \vec{0}$. First we solve the mean field equation to obtain the static profile $\vec{M}_{\mathrm{s},l}$ and then investigate the excitation modes on $\vec{M}_{\mathrm{s},l}$. In order to exclude the surface twist structure at around $l=N_{z}-1$, we use sites $l=0,\cdots, N_{z}/2-1$ for calculation of excitation spectra.  In this case, we can approximately deal with a semi-infinite system with boundary at $l=0$. The boundary condition for the diagonalization is correspondingly given by $\tilde{M}_{l=-1,N_{z}/2}^{x,y} = 0$. Note that the condition at $N_{z}/2$ gives finite size effects, but the effects on the localized mode are negligible and those on the extended mode are not very important in the following. 

\subsection{Instability modes}
We consider the same case of $D_{\perp} = 0$ as in the main text.  
In this case, instabilities are caused by a mode uniform in the plane perpendicular to the helical axis ($z$-axis), and we set $\bm{k}_{\perp} = 0$. We remark that the non-reciprocity appears only when $D_{\perp}  \neq 0$. 
Spin profiles of excited modes are shown for the basis $\vec{\tilde{e}}_{l}^{\mu} (\mu = x,y,z)$ in the following. 

\begin{figure}
\includegraphics[width = 45em]{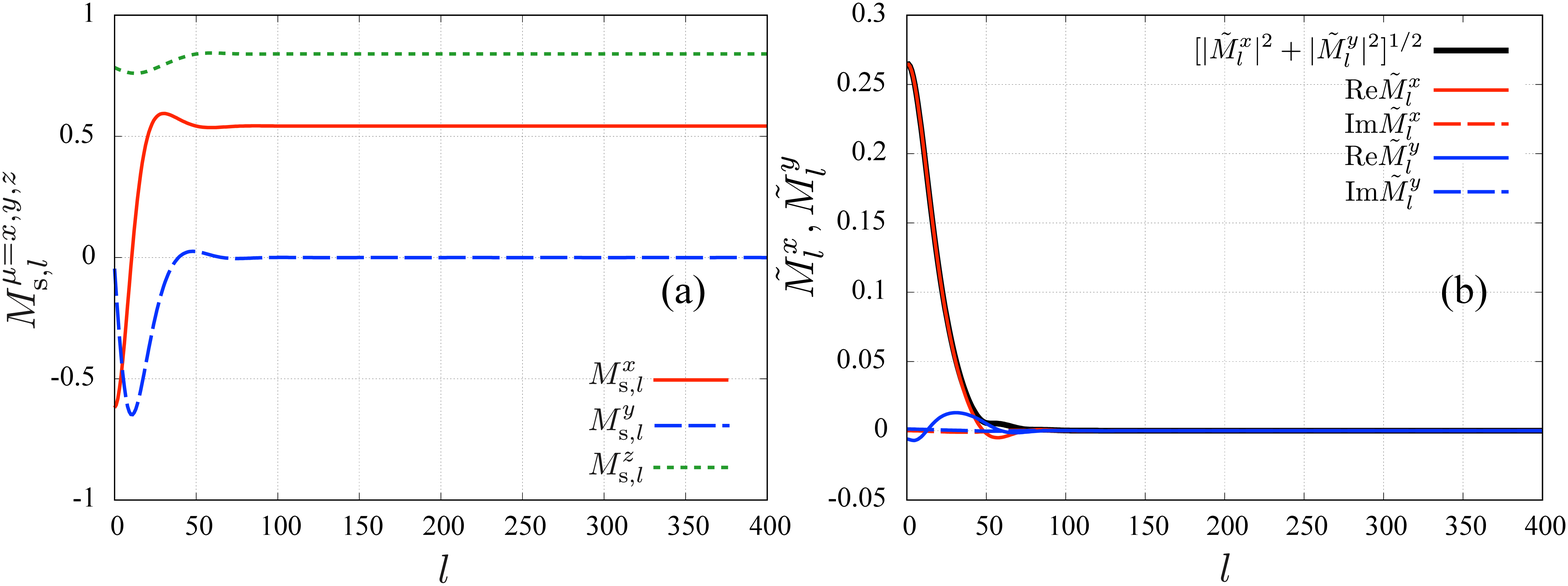}
\caption{(a) Static configurations to calculate the excitation spectra for $(H_{\mathrm{ex}}^{x}, H_{\mathrm{ex}}^{z})/H_{\dd} =(0.1474264,5.0)$. The spin profile is uniform far from the surface and a surface twist structure appears around the surface. (b) The lowest energy excitation bound to the surface. This mode causes the penetration of a soliton at the surface.}
\label{fig:static-50}
\end{figure}

\begin{figure}
\includegraphics[width = 45em]{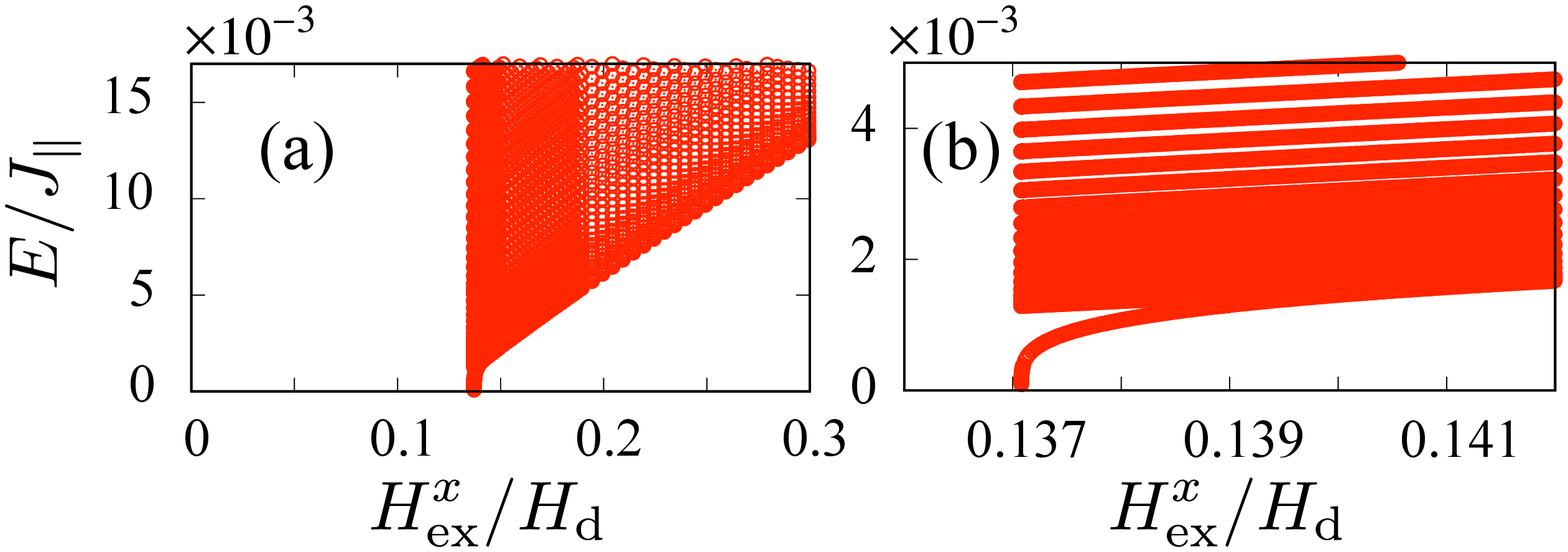}
\caption{(a) Excitation spectra for $H_{\mathrm{ex}}^{z}/H_{\dd} = 5.2$. (b) is a magnified image of (a) near the instability region. Red circles are the spectra for the static configuration given by Fig.~\ref{fig:static-52}(a). There is a low energy mode apart from the continuum spectra as well as for $H_{\mathrm{ex}}^{z}/H_{\dd} = 5.0$.}
\label{fig:E-52}
\end{figure}

\begin{figure}
\includegraphics[width = 45em]{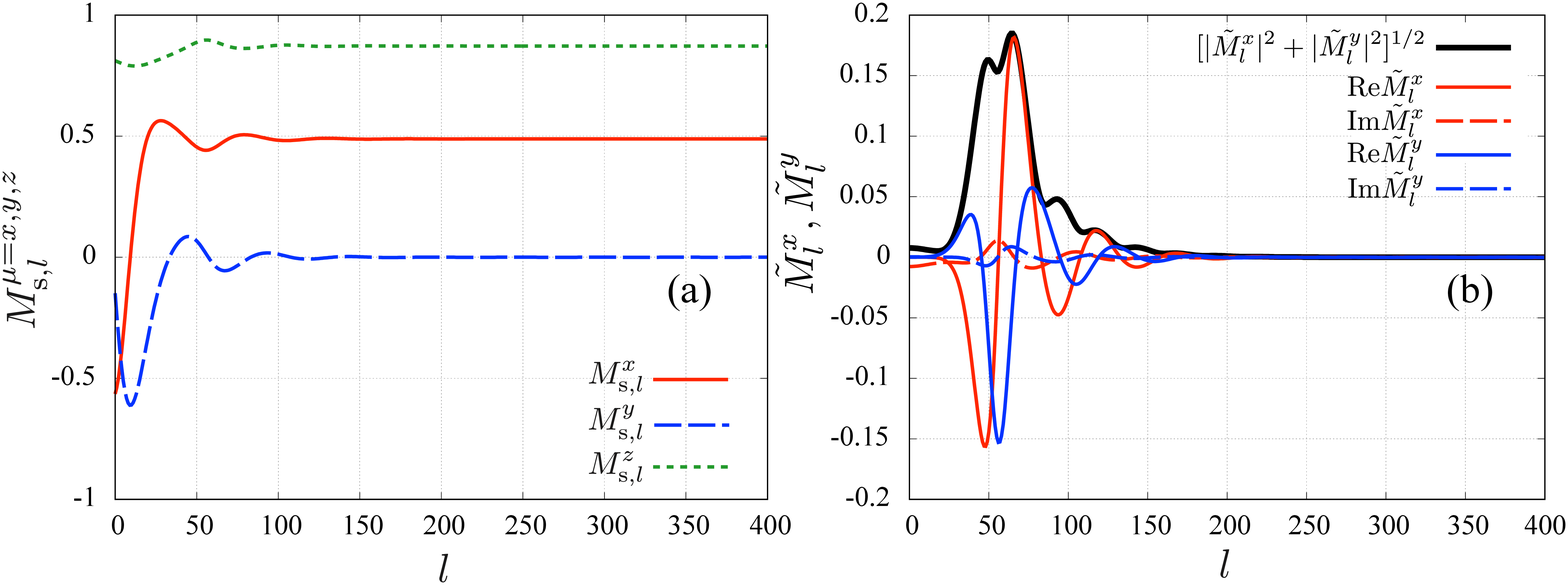}
\caption{(a) Static configurations to calculate the excitation spectra for $(H_{\mathrm{ex}}^{x}, H_{\mathrm{ex}}^{z})/H_{\dd} =(0.1370791,5.2)$, similar to Fig.~\ref{fig:static-50}(a). (b)The lowest energy excitation. This mode leads to a distorted conical order. The weight is localized around $l\sim 50$, which is about the size of the surface structure. This mode stands for the development of the oscillation in the tail of the surface structure.}
\label{fig:static-52}
\end{figure}

First we show the spin profile of an excitation mode leading to the surface instability for $H_{\mathrm{ex}}^{z}/H_{\dd} = 5.0$, discussed in the main text. We set $H_{\mathrm{ex}}^{x}/H_{\dd} =  0.1474264$. 
The spin modulation is localized around the surface, which leads to the penetration of a soliton. Then we see the excitation spectrum when we enter the PI region without crossing the barrier field $H_{\bb}$ in Fig.~\ref{fig:E-52}. We set $H_{\mathrm{ex}}^{z}/H_{\dd} = 5.2$. There is also a low energy state separated from the continuum spectra, but the weight of its wave function is away from the surface with distance about the size of the surface twist structure. The static configuration at $H_{\mathrm{ex}}^{x}/H_{\dd} = 0.1370791$ is shown in Fig.~\ref{fig:static-52}(a), and the wave function of the lowest excited state is shown in Fig.~\ref{fig:static-52}(b). This excited state is an instability mode leading to a distorted conical order. Because there is one low energy branch of the surface instability, we can expect the crossover behavior of its wave function weight between two instabilities in the vicinity of the field $\vec{H}_{\mathrm{ex}}^{*}$: The instability is a penetration of a soliton for $H_{\mathrm{ex}}^{z} < H_{\mathrm{ex}}^{*z}$, and a development of a distorted conical order for $H_{\mathrm{ex}}^{z} > H_{\mathrm{ex}}^{*z}$. However it is difficult to access this region because of enormous numerical costs. 
The instability to a distorted conical order occurs at higher field than the LA line obtained by the linear analysis. In the linear analysis, we assume the uniform state as a static configuration. In the present case, we consider the surface twist structure, which breaks the translational symmetry, and the conical order nucleates there. This is not unique to the surface structure; If there is a remnant soliton in the bulk, it becomes a nucleation point. Whether the nucleation process of the conical order occurs at the surface or an isolated soliton depends on $H_{\mathrm{ex}}^{z}$ (when we change $H_{\mathrm{ex}}^{x}$ to cause an instability). 

Finally we remark that there is another instability at higher field side when there is an isolated soliton. An isolated soliton destabilizes at some field value, and it is called the $H_{0}$ line introduced in the skyrmion system at finite temperature\cite{S-Leonov2010}. We identify this instability as the Landau instability by studying the chiral sine-Gordon model. The details about instabilities  associated with an isolated soliton are given in Ref.~\cite{S-Masaki2018}. 

%

\end{document}